\documentclass{mathincs}

\usepackage{amssymb,url}
\usepackage{comment}
\usepackage{amsthm,amsmath,amsbsy,ascmac,graphicx} 
\usepackage{eclbkbox}

 \newtheorem{thm}{Theorem}[section]
 \newtheorem{cor}[thm]{Corollary}
 \newtheorem{lem}[thm]{Lemma}
 \newtheorem{prop}[thm]{Proposition}
 \theoremstyle{definition}
 \newtheorem{defn}[thm]{Definition}
 \theoremstyle{remark}
 
 \newtheorem{ex}{Example}
 \numberwithin{equation}{section}

\DeclareMathOperator{\C}{\mathbb{C}}
\DeclareMathOperator{\N}{\mathbb{N}}
\DeclareMathOperator{\V}{\mathbb{V}}
\DeclareMathOperator{\A}{\mathbb{A}}

\DeclareMathOperator{\I}{\mathbb{I}}

\DeclareMathOperator{\hm}{hm}

\DeclareMathOperator{\hc}{hc}

\DeclareMathOperator{\Term}{Term}

\DeclareMathOperator{\het}{ht}

\def\SEN{\hrule width \textwidth}

\begin{document}
%
%
%
\title[Testing zero-dimensionality of varieties at a point]
 {Testing zero-dimensionality of varieties at a point}
\author[Katsusuke Nabeshima]{Katsusuke Nabeshima}

\address{%
Graduate School of Technology, Industrial and Social Sciences, Tokushima University, \\2-1, Minamijosanjima, Tokushima, Japan}

\email{nabeshima@tokushima-u.ac.jp}

\thanks{This work has been partly supported by JSPS  Grant-in-Aid for Scientific Research (C) (Nos 18K03214,\ 18K03320).}
\author{Shinichi Tajima}
\address{Graduate School of Science and Technology, Niigata University, \\8050, Ikarashi 2-no-cho, Nishi-ku Niigata, Japan}
\email{tajima@emeritus.niigata-u.ac.jp}



\begin{abstract}
Effective methods are introduced for testing zero-dimensionality of varieties at a point. The motivation of this paper is to compute and analyze deformations of isolated hypersurface singularities. As an application, methods for computing local dimensions are also described. For the case where a given ideal contains parameters, the proposed algorithms can output in particular a decomposition of a parameter space into strata according to the local dimension at a point of the associated varieties. The key of the proposed algorithms is the use of the notion of comprehensive Gr\"obner systems.

\end{abstract}

\maketitle
\section{Introduction}
The local dimension, the dimension of a variety at a point, is one of the most important invariants in algebraic geometry, complex analysis and singularity theory \cite{de00,Lo91,sam,whi72}. Thus, a practical tool to compute the dimension or test zero-dimensionality is required for studying local properties of a variety \cite{GLS,Nabe15,Nabe17,NaTaji18}.

In this paper, we propose two methods for testing zero-dimensionality of a variety at a point, and we generalize them to the parametric cases. The main tools of our approach are Gr\"obner bases and comprehensive Gr\"obner systems. The proposed methods do not utilize primary ideal decompositions, and are free from computation in local rings.

\begin{defn}
Let $V$ be an affine variety in $\C^n$. For $p \in V$, the dimension of $V$ at $p$, denoted $\dim_p(V)$, is the maximum dimension of an irreducible component of $V$ containing the point $p$.
\end{defn}

In singularity theory, problems that contain parameters are often studied, for instance, deformations of singularities, a family of hyperplane sections of a variety, etc. In such cases, since structures of relevant ideals or varieties may vary as parameters changes, there is a possibility that the local dimension of varieties may also depend on the values of parameters. 
We need methods to decompose a parameter space into strata according to the local dimensions of a given family of varieties.

In order to state precisely the problem, we give an example. Let $ f_0=x_1^4+x_1x_3^2+x_2^4 $ and consider $ f=f_0+t_1x_2x_3^2 $, where $ t_1 $ is a parameter.
The hypersurface defined by $ f_0 = 0 $ has an isolated singularity at the origin $ O $ in $ \C^3$, i.e., $\dim_O(\V(f_0,\frac{\partial f_0}{\partial x_1},\frac{\partial f_0}{\partial x_2},\frac{\partial f_0}{\partial x_3}))=0$. Since $ f $ has the parameter $ t_1 $, there is a possibility that the family of hypersurfaces defined by $ f=0 $ have non-isolated singularities at the origin for some values of the parameter $ t_1.$ 
In fact, if $t_1^4+1= 0$, then $f$ has a non-isolated singularity at $O$ and $\dim_O(\V(f,\frac{\partial f}{\partial x_1},\frac{\partial f}{\partial x_2},\frac{\partial f}{\partial x_3}))$ $=1$. If $t_1^4+1\neq 0$, then $f$ has an isolated singularity at $O$. We really would like the condition $t_1^4+1\neq 0$, or detect the condition $t_1^4+1= 0$ in an algorithmic manner to study local properties of the deformation of an isolated singularity.
How do we obtain such conditions? 

Basically, the condition can be obtained by testing zero-dimensionality of the variety $\V(f,$ $\frac{\partial f}{\partial x_1},\frac{\partial f}{\partial x_2},\frac{\partial f}{\partial x_3} )$ at the origin $O$. 
We show in the present paper that the methods for testing zero-dimensionality of a variety at a point can be constructed by using Gr\"obner bases. Furthermore, we generalize the methods to parametric cases by utilizing comprehensive Gr\"obner systems \cite{Kap13,Na10,N12,Nabe18,whi}. We give two different kinds of algorithms for testing zero-dimensionality at a point of a family of varieties with parameters.

Note that the resulting algorithms do not involve computation in local rings and efficiently output the necessary and sufficient conditions for zero-dimensionality. \\

This paper is organized as follows. Section~2 briefly reviews comprehensive Gr\"obner systems, and give notations that will be used in this paper. Section~3 consider the use of tangent cone and gives the discussion of the first algorithm for testing zero-dimensionality of varieties at a point. Section~4 consider the use of saturation and discusses the second algorithm for testing zero-dimensionality of varieties  at a point.  Section~5 gives results of the benchmark tests. Appendix~A gives an efficient algorithm for computing ideal quotients with parameters, that utilizes a comprehensive Gr\"obner system of a {\em module}.

\section{Preliminaries}

Let $t=\{t_1,\ldots,t_m\}$ and $x=\{x_1,\ldots, x_n\}$ be variables such that $t \cap x=\emptyset$ and $\C[t][x]$ be a polynomial ring with coefficients in a polynomial ring $\C[t]$. For $f_1,\ldots,f_s \in \C[x]$ (or $\C[t][x]$), $\langle f_1,\ldots,f_s\rangle=\{\sum_{i=1}^sh_if_i | h_1,\ldots,h_s \in \C[x] (\text{ or }\C[t][x])\}$.

A symbol $\Term(x)$ is the set of terms of $x$. Fix a term order $\succ$ on $\Term(x)$. Let $f  \in \C[x]$ (or $f \in \C[t][x]$), then, $\het(f), \hm(f)$ and $\hc(f)$ denote the head term, head monomial and head coefficient of $f$ (i.e., $\hm(f)=\het(f)\cdot \het(f)$). For $F \subset \C[x]$ (or $F \subset \C[t][x]$), $\het(F)=\{\het(f) | f\in F\}$.

For $g_1,\ldots, g_r \in \C[t]$, $\V(g_1,\ldots,g_r) \subseteq \C^m$ denotes the affine variety of $g_1,\ldots, g_r$, i.e., $\V(g_1,\ldots,g_r)=\{\bar{t}\in \C^m | g_1(\bar{t})=\cdots =g_r(\bar{t})=0\}$. We call an algebraic constructible set of a from $\V(g_1,\ldots,g_r) \backslash \V(g'_1,\ldots,g'_{r'})\subseteq \C^m$ with $g_1,\ldots,g_r, g'_1,\ldots,g'_{r'} \in \C[t]$, a stratum. Notations $\A_1, \A_2,$ $\ldots, \A_\nu$ are frequently used to represent strata.

For every $\bar{t} \in \C^m$, the canonical specialization homomorphism 
$\sigma_{\bar{t}} : \C[t][x] \rightarrow \C[x]$ 
(or $\C[t] \rightarrow \C$) is defined as the map that substitutes $t$ by $\bar{t}$ in $f(t,x) \in \C[t][x]$ (i.e., $\sigma_{\bar{t}}(f)=f(\bar{t},x) \in \C[x]$). The image $\sigma_{\bar{t}}$ of a set $F$ is denoted by $\sigma_{\bar{t}}(F)=\{\sigma_{\bar{t}}(f) | f \in F\} \subset \C[x]$. 
In this paper, the set of natural numbers $\N$ includes zero.\\

We adopt the following as a definition of a comprehensive Gr\"obner system.

\begin{defn}[comprehensive Gr\"obner system]
Let $\succ$ be a term order on $\Term(x)$.  Let $F$ be a subset of $\C[t][x]$, $\A_1, \A_2, \ldots, \A_\nu$ strata in $\C^m$ and $G_1,G_2,\ldots,G_\nu$ subsets in $\C[t][x]$. If a finite set ${\mathcal G}=\{(\A_1,G_1),(\A_2,G_2),\ldots,(\A_\nu,$ $G_\nu)\}$ of pairs satisfies the following conditions  
\begin{enumerate}
\item $\A_i \neq \emptyset$ and $\A_i \cap \A_j= \emptyset$ for $1\le i\neq j \le \nu$, 

\item for all $\bar{t} \in \A_i$, $\sigma_{\bar{t}}(G_i)$ is a minimal Gr\"obner basis of $\langle \sigma_{\bar{t}}(F) \rangle$ w.r.t. $\succ$ in $\C[x]$, and 
\item for all $\bar{t} \in \A_i$ and $f \in G_i$, $\sigma_{\bar{t}}(\hc(f))\neq 0$,
\end{enumerate}
the set ${\mathcal G}$ is called a comprehensive Gr\"obner system on $\A_1 \cup \cdots \cup \A_\ell$ for $\langle F \rangle$ w.r.t. $\succ$. We simply say that ${\mathcal G}$ is a comprehensive Gr\"obner system for $\langle F \rangle$ if $\A_1 \cup \cdots \cup \A_\ell=\C^m$. 
\end{defn} 

In several papers \cite{Kap13,Na10,N12,Nabe18}, algorithms and implementations for computing comprehensive Gr\"obner systems  are introduced. 

\begin{ex}
Let $F=\{t_1x_1x_2+x_2+1,x_1^2x_2+t_1x_1+3\}$ be a subset in $\C[t_1][x_1,x_2]$ and $\succ$ the lexicographic term order s.t. $x_1 \succ x_2$. We regard $t_1$ as a parameter in $\C$. Then, a comprehensive Gr\"obner system of $\langle F \rangle$ w.r.t. $\succ$ is \\
$\Bigl\{(\C\backslash \V(t_1^3-t_1), \{x_2^2+(2t_1^2+2)x_2-t_1^2+1,(t_1^3-t_1)x_1+x_2+3t_1^2+1\}), (\V(t_1^2-1),\{4x_1+3t_1,x_2+4\}), (\V(t_1),\{x_1^2-3,x_2+1\})\Bigl\}$.

\end{ex}

\section{Algorithm~1 (Tangent cone approach)} 

Here, we present an algorithm for testing zero-dimensionality of a variety at a point. This algorithm is based on the method described in section~9 of the famous textbook \cite{cox04}. We generalize the method to parametric cases.

Before introducing the algorithm, we prepare some notations and basic facts. 

Let $p=(p_1,\ldots,p_n)\in \C^n$, $\alpha=(\alpha_1,\ldots, \alpha_n) \in \N^n$ and $(x-p)^\alpha=(x_1-p_1)^{\alpha_1}\cdots (x_n-p_n)^{\alpha_n}$. Given any polynomial $f\in \C[x]$ of total degree $d$, $f$ can be written as a polynomial in $x_i-p_i$, namely, 
\begin{eqnarray}
f&=& f_{p,0}+f_{p,1}+\cdots + f_{p,d}\label{fpd}
\end{eqnarray} 
where $f_{p,j}$ is a linear combination of $(x-p)^\alpha$ for $\alpha_1+\cdots+\alpha_n=j \le d$. Note that $f_{p,0}=f(p)$ and $f_{p,1}=\frac{\partial f}{\partial x_1}(p)(x_1-p_1)+\cdots + \frac{\partial f}{\partial x_n}(p)(x_n-p_n)$.

The next definition is borrowed from \cite{cox04}.

\begin{defn}[tangent cone]
Let $V\subset \C^n$ be an affine variety and let $p=(p_1,\ldots,p_n)\in V$.
\begin{enumerate}
\item[(i)] If $f \in \C[x]$ is a non-zero polynomial, then $f_{p,min}$ is defined to be $f_{p,j}$, where $j$ the smallest integer such that $f_{p,j}\neq 0$ in (\ref{fpd}).\label{test}
\item[(ii)] The tangent cone of $V$ at $p$, denoted $C_p(V)$, is the variety $$C_p(V)=\V\Bigl(f_{p,min} \mid f \in \I(V)\Bigr)$$ where $\I(V)=\{f \in \C[x] | f(\bar{x})=0, \text{ for all } \bar{x} \in V \}$.
\end{enumerate}
\end{defn}

The details of the tangent cone are described in \cite{whi65,whi72}. In 1965, H. Whitney gave the following theorem.

\begin{thm}[H. Whitney \cite{whi}]\label{whit}
Let $V\subset \C^n$ be an affine variety and let $p=(p_1,\ldots,p_n)\in V$. Then, $$\dim_p(V)=\dim(C_p(V)).$$
\end{thm}

In order to compute a tangent cone, we need the following definition.

\begin{defn}
\begin{enumerate}
\item[(i)] Let $f(x) \in \C[x]$ be a polynomial of total degree $d$. Let $f(x)=\sum_{i=0}^d f_i(x)$ be the expansion of $f(x)$ as the sum of its homogeneous components where $f_i(x)$ has total degree $i$. Then, 
$$f^h(x_0,x)=\sum_{i=0}^d f_i(x)x_0^{d-i}$$
 is a homogeneous polynomial of total degree $d$ in $\C[x_0,x]$ where $x_0$ is a new variable.
\item[(ii )]Let $I$ be an ideal in $\C[x]$. We define the homogenization of $I$ to be the ideal $$I^h=\langle f^h | f \in I\rangle \subset \C[x_0,x].$$
\end{enumerate}
\end{defn}

From now on, we assume that the point $p$ is the origin $O=(0,\ldots,0)$ in $\C^n$.

\begin{prop}[Proposition~4, p.485 \cite{cox04}]\label{coxp}
Assume that the origin $O$ is a point of $V\subset \C^n$. Let $\succ$ be a block term order such that $x_0 \gg x$. Let $I$ be an ideal such that $V=\V(I)$. If $\{g_1,\ldots,g_r\}$ is a Gr\"obner basis of $I^h$ w.r.t. $\succ$, then 
$$C_O(V)=\V\Bigl(\varepsilon(g_1)_{O,min},\varepsilon(g_2)_{O,min},\ldots,\varepsilon(g_r)_{O,min}\Bigr)$$
where $\varepsilon(g_i)$ is the dehomogenization of $g_i$ for $1\le i \le r$.

\end{prop}

There exist several algorithms for computing the dimension of a variety, thus, the dimension of $C_O(V)$ can be obtained. The procedure for computing $\dim_O(V)$ is the following. \\
\begin{enumerate}
\setlength{\leftskip}{0.8cm}
\item[Step~1:] Compute $C_O(V)$.
\item[Step~2:] Compute $\dim(C_O(V))$.
\item[Return] $\dim(C_O(V))$ (as $\dim_O(V)=\dim(C_O(V))$).\\
\end{enumerate}

We  turn to the parametric cases. Let $I$ be an ideal in $\C[t][x]$ where we regard $t$ as parameters. After here we simply say that $I$ is a ``parametric'' ideal.

As described in section~2, there exist algorithms for computing comprehensive Gr\"obner systems, it  is possible to compute a comprehensive Gr\"obner system of $I^h$ w.r.t. $\succ$ in Proposition~\ref{coxp}. Therefore, Proposition~\ref{coxp} and the procedure above can be extended to the case of parametric ideals. 

The following algorithm which utilizes a comprehensive Gr\"obner system outputs a condition of zero-dimensionality of $\V(F)$ at $O$.

\noindent
\hrulefill  \vspace{0.5mm}\\
\noindent 
{\bf Algorithm~1}\vspace{-3.0mm}\\
\SEN \vspace{1mm}
\noindent 
{\bf Input:} $F=\{f_1,f_2,\ldots,f_s\} \subset \C[t][x]$ s.t. $O \in \V(F)$.\\
\hspace{1.1cm}  $\succ$: a block term order s.t. $x_0 \gg  x$ on $\Term(\{x_0\}\cup x)$. \\
{\bf Output:} $\A\subset \C^m$: For all $\bar{t}\in \A$, $\dim_{O}(\V(\sigma_{\bar{t}}(F)))=0$ (i.e., $\V(\sigma_{\bar{t}}(F))$ has an isolated point at $O$). For all $\bar{t}\in \C^m\backslash \A$, $\dim_{O}(\V(\sigma_{\bar{t}}(F)))\neq 0$.\\
{\bf BEGIN} \\
$\A \gets \emptyset$;\\
${\mathcal G}\gets $ Compute a comprehensive Gr\"obner system of $\langle f_1^h,f_2^h, \ldots, f_s^h \rangle$ w.r.t. $\succ$ in $\C[t][x_0,x]$;\\
{\bf while} ${\mathcal G} \neq \emptyset$ {\bf do}\\
 \ \ \ \ Select $(\A',G')$ from ${\mathcal G}$; ${\mathcal G}\gets {\mathcal G}\backslash \{(\A',G')\}$;\\
 \ \ \ \ $M\gets \het(G')$ w.r.t. $\succ$; \\
 \ \ \ \ $C_O\gets \{\varepsilon(h)| h \in M\}$ in $\C[x]$; \\
 \ \ \ \ \ \ {\bf if} $\dim(C_O)=0$ {\bf then}\\
 \ \ \ \ \ \ \ \ \ \ $\A \gets \A' \cup \A$;\\
 \ \ \ \ \ \ {\bf end-if}\\
{\bf end-while}\\
{\bf return} $\A$;\\
{\bf END} \vspace{-2.0mm}  \\
\hrulefill \vspace{-1.5mm} \\

Since the algorithms \cite{Kap13,Na10,N12,Nabe18,whi} for computing comprehensive Gr\"obner systems always terminate and return a finite set of pairs, Algorithm~1 also terminates. 
The correctness follows from Theorem~\ref{whit} and Proposition~\ref{coxp}.

Note that Algorithm~1 contains a part of computing local dimensions. Thus, it can be naturally generalized to a method for decomposing a parameter space into strata according to the local dimensions of a given family of varieties. \\

We illustrate Algorithm~1 with the following example.

\begin{ex}
Let $f=x_1^3+t_1x_1^2x_2^4+x_2^{12} \in \C[t_1][x_1,x_2]$, $F=\{f,\frac{\partial f}{\partial x_1}, \frac{\partial f}{\partial x_2}\}$ and $I=\langle F \rangle$ where $t_1$ is a parameter. Let $\succ$ be the total degree lexicographic term order with $x_1 \succ x_2$.

A comprehensive Gr\"obner systems of 
$$
I^h=\langle x_1^3x_0^9+t_1x_1^2x_2^4x_0^6+x_2^{12},3x_1^2x_0^2+2t_1x_2^4,4t_1x_1^2x_2^3x_0^6+12x_2^{11} \rangle 
$$  
w.r.t. the block term order with $x_0 \gg \{x_1, x_2\}$, in $\C[t_1][x_0,x_1,x_2]$, is
\begin{eqnarray*}
\{(\C\backslash \V(t_1(4t_1^3+27)), \{(4t_1^3+27)x_1x_2^{11},(4t_1^3+27)x_2^{15},3x_0^3x_1^2+2t_1x_1x_2^4, \\
2t_1^2x_0^3x_1x_2^7-9x_2^{11}\}),\\
(\V(4t_1^3+27), \{3x_0^3x_1^2+2t_1x_1x_2^4,3x_0^3x_1x_2^7+2t_1x_2^{11}\}),(\V(t_1), \{x_2^{11},x_0x_1^2\})\}.
\end{eqnarray*} 
Hence, 
\begin{enumerate}
\setlength{\leftskip}{-0.3cm}
\item[$\cdot$] if $t_1$ belongs to $\C\backslash \V(t_1(4t_1^3+27)$, then $C_O(\V(I))=\V(x_1x_2^{11},x_2^{15},x_1^{2},x_1^2x_2^7)$ and $\dim_O(\V(I))$ $=0$, 
\item[$\cdot$] if $t_1$ belongs to $\V(4t_1^3+27)$, then $C_O(\V(I))=\V(x_1^2,x_1x_2^{7})$ and $\dim_O(\V(I))=1$, 
\item[$\cdot$] if $t_1$ belongs to $\C\backslash \V(t_1)$, then $C_O(\V(I))=\V(x_1^2,x_2^{11})$ and $\dim_O(\V(I))=0$.
\end{enumerate}
Therefore, for all $\bar{t}\in \C \backslash \V(4t_1^3+27)$, $\dim_O(\V(\sigma_{\bar{t}}(I)))=0$, namely, $f$ has an isolated singularity at the origin $O$.

\end{ex}

\section{Algorithm~2 (Saturation approach)}

in this section, we consider the use of saturation and introduce an alternative method for testing zero-dimensionality of a variety at a point. We present an algorithm for testing zero-dimensionality of a family of varieties at a point, that utilizes a comprehensive Gr\"obner system of a {\it module}. Furthermore, we improve the algorithm in speed and give an efficient algorithm. We also show that, according to the concept of Chevalley dimension, local dimensions of varieties can also be computed by utilizing saturation.

\subsection{Saturation approach}
Let $I, J$ be ideals in $\C[x]$. The ideal quotient of $I$ by $J$ is $I : J =\{h \in \C[x] | hg \in I \text{ for all } g \in J\}.$ 
The saturation of $I$ with respect to $J$ is the ideal  
$$
I : J^\infty=\{h \in \C[x] | h J^r \subset I \text{ for some } r >0\}.
$$
The saturation $I : J^\infty$ is the ideal at which the chain
$$
I : J \subseteq I : J^2 \subseteq I : J^3\subseteq \cdots 
$$
stabilizes.

Now, we give the following main theorem which is utilized to construct the new algorithm for testing the zero-dimensionality of a variety at a point.

\begin{thm}\label{main}
Let $F \subset \C[x]$, ${\mathfrak m}=\langle x_1,\ldots,x_n\rangle \subset \C[x]$ and $O \in \V(F)$. Let $G$ be a basis of the ideal $\langle F \rangle : {\mathfrak m}^\infty$ in $\C[x]$. Then, the affine variety $\V(F)$ has an isolated point at the origin $O$ if and only if there exists $g \in G$ such that $g(O)\neq 0$ (i.e., $g$ has non-zero constant term).

\end{thm}
\begin{proof}
As $G$ is a basis of $\langle F \rangle : {\mathfrak m}^\infty$, $\V(G)=\V(\langle F \rangle : {\mathfrak m}^\infty)$ is the Zariski closure $\overline{\V(F)\backslash \{O\}}$. The variety $\V(F)$ can be written as $\V(F)=V_1\cup V_2 \cup \cdots \cup V_\nu$ (finite union) where $V_1, V_2,\ldots, V_\nu$ are distinct irreducible varieties. 

First, assume that the affine variety $\V(F)$ has an isolated point at the origin $O$. Then, one of $V_i$ must be $\{O\}$ and other varieties does not contain $O$ where $i \in \{1,2,\ldots,\nu\}$. Without loss of generality, set $V_1=\{O\}$. Then, 
$$\V(G)=\V(\langle F \rangle : {\mathfrak m}^\infty)=\overline{\V(F)\backslash \{O\}}=V_2 \cup \cdots \cup V_\nu.$$
As $O \notin V_2 \cup \cdots \cup V_\nu=\V(G)$, there exists $g \in G$ such that $g(O)\neq 0$.

Next, assume that there exists $g \in G$ such that $g(O)\neq 0$. Since $O \in \V(F)$, 
$$\V(G)= \V(\langle F \rangle : {\mathfrak m}^\infty)=\overline{\V(F)\backslash \{O\}}\text{ and }O \notin \V(G),$$ 
there exists an irreducible variety $\{O\}$ in $\V(F)$. Therefore, $\V(F)$ has an isolated point at the origin $O$.
\end{proof}

The following corollary is a direct consequence of Theorem~\ref{main}.
\begin{cor}
Let $f \in \C[x]$, ${\mathfrak m}=\langle x_1,\ldots,x_n\rangle \subset \C[x]$ and $O \in \V(f,\frac{\partial f}{\partial x_1},\ldots,\frac{\partial f}{\partial x_n})$. Let $G$ be a basis of $\langle \frac{\partial f}{\partial x_1},\ldots,\frac{\partial f}{\partial x_n}\rangle : {\mathfrak m}^\infty$ (or $\langle f, \frac{\partial f}{\partial x_1},\ldots,\frac{\partial f}{\partial x_n}\rangle : {\mathfrak m}^\infty$). Then, the hypersurface, defined by $f$, has an isolated singularity at $O$ if and only if there exists $g \in G$ such that $g(O)\neq 0$.
\end{cor}

\begin{ex}
Let us consider $f_1=x_1^2x_3+x_2x_3^2+x_2^5+x_2^3x_3, f_2=x_1^2x_3+x_2x_3^2+x_2^5+2x_2^3x_3$ and ${\mathfrak m}=\langle x_1,x_2,x_3\rangle$ in $\C[x_1,x_2,x_3]$. Let $\succ$ be the total degree lexicographic term order with $x_1 \succ x_2 \succ x_3$. Then, the reduced Gr\"obner basis of $\langle \frac{\partial f_1}{\partial x_1}, \frac{\partial f_1}{\partial x_2}, \frac{\partial f_1}{\partial x_3}\rangle : {\mathfrak m}^\infty$ w.r.t. $\succ$ is 
$$\{1\},$$ 
and  the reduced Gr\"obner basis of $\langle \frac{\partial f_2}{\partial x_1}, \frac{\partial f_2}{\partial x_2}, \frac{\partial f_2}{\partial x_3} \rangle : {\mathfrak m}^\infty$ w.r.t. $\succ$ is 
$$\{x_2^2+x_3,x_1\}.$$

Therefore, $f_1$ has an isolated singularity at $O$, and $f_2$ does not have an isolated singularity at $O$.
\end{ex}

We turn to the parametric cases. There exists an algorithm for computing a comprehensive Gr\"obner system of the saturation of $\langle F \rangle$ w.r.t. a given parametric ideal. The algorithm is given in Appendix~A. Therefore, Theorem~\ref{main} is generalized to the parametric cases. 

\noindent
\hrulefill  \vspace{0.5mm}\\
\noindent 
{\bf Algorithm~2-1}\vspace{-3.0mm}\\
\SEN \vspace{1mm}
\noindent 
{\bf Input:} $F=\{f_1,f_2,\ldots,f_s\} \subset \C[t][x]$ s.t. $O \in \V(F)$,  ${\mathfrak m}=\langle x_1,\ldots,x_n\rangle$. \\
\hspace{1.1cm} $\succ$: a term order on $\Term(x)$.  \\
{\bf Output:} $\A\subset \C^m$: For all $\bar{t}\in \A$, $\dim_{O}(\V(\sigma_{\bar{t}}(F)))=0$ (i.e., $\V(\sigma_{\bar{t}}(F))$ has an isolated point at $O$). For all $\bar{t}\in \C^m\backslash \A$, $\dim_{O}(\V(\sigma_{\bar{t}}(F)))\neq 0$. \\
{\bf BEGIN} \\
$\A \gets \emptyset$;\\
${\mathcal G}\gets$ Compute a comprehensive Gr\"obner system of $\langle F \rangle : {\mathfrak m}^\infty$ w.r.t. $\succ$;\\
{\bf while} ${\mathcal G}\neq \emptyset$ {\bf do}\\
 \ \ \ \ Select $(\A',G')$ from ${\mathcal G}$; ${\mathcal G}\gets {\mathcal G}\backslash \{(\A',G')\}$;\\
 \ \ \ \ \ \ {\bf if} $\exists g \in G'$ s.t. $g(O)\neq 0$ {\bf then}\\
 \ \ \ \ \ \ \ \ \ \ $\A \gets \A' \cup \A$;\\
 \ \ \ \ \ \ {\bf end-if}\\
{\bf end-while}\\
{\bf return} $\A$;\\
{\bf END} \vspace{-2.0mm}  \\
\hrulefill \vspace{-1.5mm} \\

\noindent

The correctness and termination of Algorithm~2-1 follows from Theorem~\ref{main} and that of the algorithm for computing comprehensive Gr\"obner systems.\\

We illustrate Algorithm~1 with the following example.

\begin{ex}
Let $f=x_1^3+x_1x_3^2+t_1x_1x_2^3+x_2^3x_3 \in \C[t_1][x_1,x_2,x_3]$ and $\succ$ the total degree reverse lexicographic term order with the coordinate $x_1 \succ x_2 \succ x_3$. 

A comprehensive Gr\"obner system of $\langle f, \frac{\partial f}{\partial x_1},\frac{\partial f}{\partial x_2}, \frac{\partial f}{\partial x_3}\rangle : \langle x_1,x_2,x_3 \rangle^\infty$ w.r.t. $\succ$ is 
$$
\{(\C\backslash \V(t_1^2+1), \{1\}),(\V(t_1^2+1), \{x_1-t_1x_3,x_2^3+2t_1x_3^2\})\}.$$
Hence, 
\begin{enumerate}
\item[$\cdot$] if $t_1$ belongs to $\C\backslash \V(t_1^2+1)$, then $\dim_O\Bigl(\V(f, \frac{\partial f}{\partial x_1},\frac{\partial f}{\partial x_2}, \frac{\partial f}{\partial x_3})\Bigr)=0$, 
\item[$\cdot$] if $t_1$ belongs to $\V(t_1^2+1)$, then $\dim_O\Bigl(\V(f, \frac{\partial f}{\partial x_1},\frac{\partial f}{\partial x_2}, \frac{\partial f}{\partial x_3})\Bigr)\neq 0$.
\end{enumerate}
Therefore, for all $\bar{t}\in \C \backslash \V(t_1^2+1)$, $\sigma_{\bar{t}}(f)$ has an isolated singularity at the origin $O$.\\
\end{ex}

\subsection{Improvement}

We improve Algorithm~2-1 in computation speed. The following lemma alows us to devise an efficient and practical algorithm for computing the saturation $\langle F \rangle : {\mathfrak m}^\infty$.

\begin{lem}\label{imp}
Let $F \subset \C[x]$ and ${\mathfrak m}=\langle x_1,\ldots,x_n\rangle$. For all $\alpha_1,\alpha_2,\ldots,\alpha_n \in \N \backslash \{0\}$, 
$$\Bigl(\langle F \rangle : \langle x_1^{\alpha_1}, x_2^{\alpha_2}, \ldots, x_n^{\alpha_n}\rangle \Bigr) : {\mathfrak m}^\infty=\langle F \rangle : {\mathfrak m}^\infty.$$
\end{lem}
\begin{proof}
There exists $\beta \in \N$ such that 
$$\langle F \rangle : {\mathfrak m}^{\beta-1}\subset \langle F \rangle : {\mathfrak m}^\beta=\langle F \rangle : {\mathfrak m}^{\beta+1}=\cdots.$$
Let $J=\langle x_1^{\alpha_1}, x_2^{\alpha_2}, \ldots, x_n^{\alpha_n}\rangle$ and $\alpha = \max\Bigl(\{\alpha_1,\alpha_2,\ldots,\alpha_n\}\Bigr)$. Obviously, $$\Bigl(\langle F \rangle : J \Bigr) : {\mathfrak m}^\beta= \langle F \rangle  : J\cdot{\mathfrak m}^\beta.$$ 
Since $J\cdot {\mathfrak m}^\beta \subseteq {\mathfrak m}^\beta$, thus $\langle F \rangle : {\mathfrak m}^{\beta}\subseteq \langle F \rangle : J\cdot {\mathfrak m}^\beta$. Take a sufficiently large number $N$ such that $N >\alpha + \beta$, then ${\mathfrak m}^N \subseteq J\cdot {\mathfrak m}^\beta$. Hence, 
$$
\langle F \rangle : {\mathfrak m}^{\beta}\subseteq \langle F \rangle : J\cdot {\mathfrak m}^\beta \subseteq \langle F \rangle : {\mathfrak m}^N.
$$
As $\langle F \rangle : {\mathfrak m}^{\beta}=\langle F \rangle : {\mathfrak m}^N$, we have $\langle F \rangle : {\mathfrak m}^\beta = \langle F \rangle : J\cdot {\mathfrak m}^{\beta}$. Therefore, 
$\Bigl(\langle F \rangle : \langle x_1^{\alpha_1}, \ldots, x_n^{\alpha_n}\rangle \Bigr) : {\mathfrak m}^\infty=\langle F \rangle : {\mathfrak m}^\infty.$
\end{proof}

The lemma above leads the following procedure for computing $\langle F \rangle : {\mathfrak m}^\infty$.

\begin{enumerate}
\setlength{\leftskip}{0.8cm}
\item[Step~1:] Compute a basis $G$ of $\langle F \rangle : \langle x_1^{\alpha_1}, x_2^{\alpha_2}, \ldots, x_n^{\alpha_n}\rangle$.
\item[Step~2:] Compute a basis $G'$ of $\langle G \rangle : {\mathfrak m}^\infty$.
\item[Return] $G'$.\\
\end{enumerate}

Notice that in the procedure above arbitrary positive integers $ \alpha_1,\ldots,\alpha_n $ can be used to compute $\langle F \rangle : {\mathfrak m}^\infty$.
 Our strategy of choosing the integers is the following. \\

Let $\displaystyle f=\sum_{{\bf r} \in \N^n}a_{{\bf r}}x^{{\bf r}}$ be a non-zero polynomial in $\C[x]$ and $F \subset \C[x]$. We set 
$$\text{mdeg}_{x_i}(f):=\max\Bigl(\{\gamma_i | \ {\bf r}=(\gamma_1,\ldots,\gamma_i,\ldots,\gamma_n) \in \N^n, a_{{\bf r}}\neq 0\}\Bigr),$$
and $\text{mdeg}_{x_i}(F):=\max(\{\text{mdeg}_{x_i}(g)| g \in F\})$ where $i \in \{1,\ldots,n\}$. 
In Algorithm~{2-2}, we take 
$$\alpha=\max\Bigl(\{\text{mdeg}_{x_1}(F),\text{mdeg}_{x_2}(F),\ldots,\text{mdeg}_{x_n}(F)\}\Bigr)$$
as $\alpha_1,\ldots,\alpha_n$ to compute a basis of $\langle F \rangle : {\mathfrak m}^\infty$, namely, $\alpha_1=\alpha_2=\cdots=\alpha_n=\alpha$.\\

Lemma~\ref{imp} and the strategy above yield the following improvement.\\
 \ \\
\noindent
\hrulefill  \vspace{0.5mm}\\
\noindent 
{\bf Algorithm~2-2}\vspace{-3.0mm}\\
\SEN \vspace{1mm}
\noindent 
{\bf Input:} $F=\{f_1,f_2,\ldots,f_s\} \subset \C[t][x]$ s.t. $O \in \V(F)$,  ${\mathfrak m}=\langle x_1,\ldots,x_n\rangle$.\\
\hspace{1.1cm}  $\succ$: a term order on $\Term(x)$.  \\
{\bf Output:} $\A^\ast \subset \C^m$: For all $\bar{t}\in \A^\ast$, $\dim_{O}(\V(\sigma_{\bar{t}}(F)))=0$ (i.e., $\V(\sigma_{\bar{t}}(F))$ has an isolated point at $O$). For all $\bar{t}\in \C^m\backslash \A^\ast$, $\dim_{O}(\V(\sigma_{\bar{t}}(F)))\neq 0$. \\
{\bf BEGIN} \\
$\A^\ast \gets \emptyset$; \ $\alpha \gets\max\Bigl(\{\text{mdeg}_{x_1}(F),\text{mdeg}_{x_2}(F),\ldots,\text{mdeg}_{x_n}(F)\}\Bigr)$; \\
${\mathcal G}\gets$ Compute a comprehensive Gr\"obner system of $\langle F \rangle : \langle x_1^\alpha,x_2^\alpha,\ldots, x_n^\alpha\rangle$ w.r.t. $\succ$;\\
{\bf while} ${\mathcal G}\neq \emptyset$ {\bf do}\\
 \ \ Select $(\A,G)$ from ${\mathcal G}$; ${\mathcal G}\gets {\mathcal G}\backslash \{(\A,G)\}$;\\
 \ \ ${\mathcal G}'\gets$ Compute a comprehensive Gr\"obner system of $\langle G \rangle : {\mathfrak m}^\infty$ w.r.t. $\succ$ on $\A$;\\
 \ \ \ \ {\bf while} ${\mathcal G}'\neq \emptyset$ {\bf do}\\
 \ \ \ \ \ \ \ \ Select $(\A',G')$ from ${\mathcal G}'$; ${\mathcal G}'\gets {\mathcal G}\backslash \{(\A',G')\}$;\\
 \ \ \ \ \ \ \ \ \ \ \ \ {\bf if} $\exists g \in G_2$ s.t. $g(O)\neq 0$ {\bf then}\\
 \ \ \ \ \ \ \ \ \ \  \ \ \ \ $\A^\ast \gets \A' \cup \A^\ast$;\\
 \ \ \ \ \ \ \ \ \ \ \ \ {\bf end-if}\\
 \ \ \ \ {\bf end-while}\\
{\bf end-while}\\
{\bf return} $\A^\ast$;\\
{\bf END} \vspace{-2.0mm}  \\
\hrulefill \vspace{-1.5mm} \\

We have implemented Algorithm~2-2 in the computer algebra system {\sf Risa/Asir} \cite{NT92}. We give some outputs of our implementation in the following examples.

\begin{ex}\label{ex5}
Let $f=x_1^3x_2+t_1x_1^2x_2^4+x_2^{10}+t_2x_2^{11} \in \C[t_1,t_2][x_1,x_2]$ and $V=\V(\frac{\partial f}{\partial x_1},\frac{\partial f}{\partial x_2})$. Our implementation outputs the following.
\begin{enumerate}
\setlength{\leftskip}{-0.3cm}
\item[$\cdot$] If $(t_1,t_2)$ belongs to $W=\C^2\backslash \V(4t_1^3+27,t_2),$ then $\dim_O(V)=0$, namely, the hypersurface $S$, defined by $f$, has an isolated singularity at $O$. 
\item[$\cdot$] If $(t_1,t_2)$ does not belong to $W$, then $\dim_O(V)\neq 0$. The hypersurface $S$ does not have an isolated singularity at $O$.
\end{enumerate}
\end{ex}

\begin{ex}
Let $f=x_1x_3^2+x_1^4+x_2^4+t_1x_2x_3^2+t_2x_1^2x_2^2 \in \C[t_1,t_2][x_1,x_2,x_3]$ and $V=\V(\frac{\partial f}{\partial x_1},\frac{\partial f}{\partial x_2},\frac{\partial f}{\partial x_3})$. Our implementation outputs the following.
\begin{enumerate}
\setlength{\leftskip}{-0.3cm}
\item[$\cdot$] If $(t_1,t_2)$ belongs to $$W=\Bigl(\C^2\backslash \V((t_1^4+t_1^2t_2+1)(t_2-2)(t_2+2)\Bigr) \cup \Bigl(\V(t_2)\backslash \V(t_1^4+1,t_2)\Bigr),$$ then $\dim_O(V)=0$, namely, the hypersurface $S$, defined by $f$, has an isolated singularity at $O$. 
\item[$\cdot$] If $(t_1,t_2)$ does not belong to $W$, then $\dim_O(V)\neq 0$. The hypersurface $S$ does not have an isolated singularity at $O$.
\end{enumerate}

\end{ex}

We will see in section~5 how the use of $\langle F \rangle : \langle x_1^\alpha,x_2^\alpha,\ldots, x_n^\alpha\rangle$ reduces the cost of computation of the saturation $\langle F \rangle : {\mathfrak m}^\infty$ drastically.\\

\subsection{Primary ideal component at $O$}

Let $\langle F \rangle$ be the ideal generated by $F$ s.t. $O \in \V(F)$. Let 
$$\langle F \rangle=Q_0 \cap Q_1 \cap \ldots \cap Q_{\nu}$$
be the primary ideal decomposition of the ideal $\langle F \rangle$. Let $S$ denote the saturation $\langle F \rangle : {\mathfrak m}^\infty$ where ${\mathfrak m}=\langle x_1,\ldots, x_n\rangle$ is the maximal ideal in $\C[x]$.

Assume that $O \notin \V(S)$. Then, we have 
$$\langle F \rangle =Q_0\cap S\text{ \ \ and  \ \ } Q_0 =\langle F \rangle : S.$$
Therefore,  the primary component $Q_0$ at $O$ of $\langle F \rangle$ such that $\V(Q_0)=\{O\}$ can also be computed by using the saturation $S=\langle F \rangle : {\mathfrak m}^\infty$. 

The method above works for parametric cases, too.

\begin{ex}
Let $f=x_1^3x_2+x_1^2x_2^4+t_1x_2^{10} \in \C[t_1][x_1,x_2]$ and $F=\{\frac{\partial f}{\partial x_1},\frac{\partial f}{\partial x_2}\}$. Let $\succ$ be the total degree lexicographic term order with $x_1 \succ x_2$. A comprehensive Gr\"obner system of $\langle F  \rangle : {\mathfrak m}^\infty$ w.r.t. $\succ$ is 
\begin{eqnarray*}
\{(\V(t_1),\{1\}),(\C\backslash \V(t_1),\{2910897t_1^3x_1+16385050t_1x_2+14895500, \\
-9801t_1^2x_1+52855t_1x_2^2+48050x_2\})\}.
\end{eqnarray*}
\begin{enumerate}
\item[(i)] A comprehensive Gr\"obner system of $\langle F  \rangle : \langle 1 \rangle$ on $\V(t_1)$ w.r.t. $\succ$ is  
$$\{(\V(t_1), Q_0')\},$$
where $Q'_0=\{3x_1^2x_2+2x_1x_2^4,x_1^4x_2,x_1^5,2x_1^4+53x_1^3x_2^3,x_1^3+4x_1^2x_2^3+10x_2^9\}$.
This means that if $t_1=0$, then $\langle Q_0'\rangle$ is the primary ideal component such that $\V(Q_0')=\{O\}$.

\item[(ii)] A comprehensive Gr\"obner system of 
\begin{center}
$\langle F  \rangle : \langle 2910897t_1^3x_1+16385050t_1x_2+14895500, -9801t_1^2x_1+52855t_1x_2^2+48050x_2 \rangle$
\end{center}
 on $\C\backslash \V(t_1)$ w.r.t. $\succ$ is 
\begin{center}
$\{ (\C\backslash \V(t_1), Q_0'')\}$
\end{center}
where $Q_0''=\{3x_1^2x_2+2x_1x_2^4,x_1^4x_2,x_1^5,2x_1^4+53x_1^3x_2^3,-1331t_1^3x_1^4+(-32065t_1^2x_2^2+29150t_1$ $x_2+5300)x_1^3+21200x_1^2x_2^3+53000x_2^9\}$. Thus, if $t_1\neq 0$, $\langle Q_0''\rangle$ is the primary ideal component such that $\V(Q_0')=\{O\}$.\\
\end{enumerate}

\end{ex}

\subsection{Local dimensions}
In this subsection, we give an algorithm for computing the local dimension at $O$. 

Let $F$ be a set of polynomials in $\C[x]$. Let $E$ be a set of families of linear polynomials 
\begin{eqnarray*}
\begin{array}{c}
x_1+u_{11}x_{\ell+1}+u_{12}x_{\ell+2}+\cdots +u_{1 n-\ell}x_n, \\
x_2 +u_{21}x_{\ell+1}+u_{22}x_{\ell+2}+\cdots +u_{2 n-\ell}x_n,  \\
x_3 +u_{31}x_{\ell+1}+u_{32}x_{\ell+2}+\cdots +u_{3 n-\ell}x_n,         \\
      \vdots             \\
x_{\ell} + u_{\ell 1}x_{\ell+1}+u_{\ell 2}x_{\ell+2}+\cdots +u_{\ell n-\ell}x_n,\\ 
\ \\
\end{array} 
 \end{eqnarray*}  
\noindent 
in $\C(u)[x]$ where $u=\{u_{11},\ldots,u_{1 n-\ell},u_{21},\ldots,u_{2\ell-n},\ldots,u_{\ell 1}, \ldots, u_{\ell n-\ell}\}$, $\ell \le  n$ and $\C(u)$ is the fields of rational functions with $u$. Then, we have the following proposition.

\begin{prop}\label{prodim}
Let ${\mathfrak m}=\langle x_1,\ldots,x_n\rangle \subset \C[x]$ and $O \in \V(F)$. Let $G$ be a basis of the ideal $\langle F \cup E \rangle : {\mathfrak m}^\infty$ in $\C(u)[x]$. Let  $\ell$ be the minimum number that satisfies 
\begin{center}
``there exists $g \in G$ such that $g(O)\neq 0$.''
\end{center}
 Then, $\dim_{O}(\V(F))= \ell$.
\end{prop}
\begin{proof}
By Theorem~\ref{main}, if there exists $g \in G$ such that $g(O)\neq 0$, then $\dim_{O}(\V(F \cup E))=\dim_{O}(\V(F)\cap \V(E))=0$.  Note that $\C(u)$ is the fields of rational functions and $E$ is a set of $\ell$ linear polynomials with $O \in \V(E)$ Hence, $\dim_{O}(\V(E))=n-\ell$. Since $\ell$ is the minimum number, it follows from the classical action lemma or the concept of Chevally dimension that  $\dim_{O}(\V(F))= \ell$.
\end{proof}

By this proposition, we can construct an algorithm for computing $\dim_{O}(\V(F))$ as follows.

\noindent
\hrulefill  \vspace{0.5mm}\\
\noindent 
{\bf Algorithm~2-3}\vspace{-3.0mm}\\
\SEN \vspace{1mm}
\noindent 
{\bf Input:} $F=\{f_1,f_2,\ldots,f_s\} \subset \C[x]$ s.t. $O \in \V(F)$,  ${\mathfrak m}=\langle x_1,\ldots,x_n\rangle$, \\
\hspace{1.1cm}  $\succ$: a term order on $\Term(x)$.  \\
{\bf Output:} $\dim_O(\V(F))$.\\
{\bf BEGIN} \\
$\ell \gets 0$; \ $\text{flag} \gets 0$; \ $E\gets \emptyset$; \ $U \gets \emptyset$; \\
{\bf while} $\text{flag} \neq 1$ {\bf do} \\
 \ \ \ $G \gets$ Compute a basis of $\langle F \cup E\rangle : {\mathfrak m}^\infty$ w.r.t. $\succ$ in $\C(U)[x]$; \ \ \verb|/*if |$U=\emptyset$, \verb|then| $\C(U)=\C$.\verb|*/| \\
 \ \ \ \  \ \ \ \ {\bf if} $\exists g \in G$ s.t. $g(O)\neq 0$ {\bf then }\\
 \ \ \ \  \ \ \ \ \ \ \  \ \ \ \ $\text{flag} \gets 1$;\\
 \ \ \ \  \ \ \ \ {\bf else}\\
 \ \ \ \  \ \ \ \  \ \ \ \ \ \ \ $\ell \gets \ell +1$; \\ 
 \ \ \ \  \ \ \ \  \ \ \ \ \ \ \ $E \gets \{x_1+u_{11}x_{\ell+1}+\cdots +u_{1 n-\ell}x_n, x_2+u_{21}x_{\ell+1}+\cdots +u_{2 n-\ell}x_n,$\\
\hspace{8cm} $\cdots, x_{\ell} + u_{\ell 1}x_{\ell+1}+\cdots +u_{\ell n-\ell}x_n\}$;\\ 
 \ \ \ \  \ \ \ \  \ \ \ \ \ \ \ $U \gets \{u_{11},\ldots,u_{1 n-\ell},u_{21},\ldots,u_{2\ell-n},\ldots,u_{\ell 1}, \ldots, u_{\ell n-\ell}\}$;\\
 \ \ \ \  \ \ \ \ {{\bf end-if}\\
{\bf end-while}\\
{\bf return} $\ell$;\\
{\bf END} \vspace{-2.0mm}  \\
\hrulefill \vspace{-1.5mm} \\

We illustrate Algorithm~2-3 with the following examples.

\begin{ex}
Let $f=x_1^4+x_2^6+2x_1^2x_2^3 \in \C[x_1,x_2]$ and $F=\{\frac{\partial f}{\partial x_1},\frac{\partial f}{\partial x_2}\}$. Let $\succ$ be the total degree reverse lexicographic term order with $x_1 \succ x_2$. 

The reduced Gr\"obner basis of $\langle F  \rangle : \langle x_1,x_2\rangle^\infty$ w.r.t. $\succ$, in $\C[x_1,x_2]$, is 
$$\{x_1^2+x_2^3\}.$$ 
Thus, $\dim_O(\V(F))\neq 0$. 

Next, let us consider the case $\ell =1$. Let $E=\{x_1+u_{11}x_2\}$. The reduced Gr\"obner basis of $\langle F  \cup E \rangle : \langle x_1,x_2\rangle^\infty$ w.r.t. $\succ$ is, $\C(u_{11})[x_1,x_2]$ is 
$$\{x_2+u_{11}^2,x_1-u_{11}^3\}.$$ 
Hence, as $\dim_O(\V(F\cup E))=0$, we obtain $\dim_{O}(\V(F))=1$. 

\end{ex}

\begin{ex}
Let $f=x_1^3+x_2x_3^2+2x_1^2x_2^2+x_1x_2^4 \in \C[x_1,x_2,x_3]$ and $F=\{\frac{\partial f}{\partial x_1},\frac{\partial f}{\partial x_2},\frac{\partial f}{\partial x_3}\}$. Let $\succ$ be the total degree reverse lexicographic term order with $x_1 \succ x_2 \succ x_3$. 

The reduced Gr\"obner basis of $\langle F  \rangle : \langle x_1,x_2,x_3\rangle^\infty$ w.r.t. $\succ$, in $\C[x_1,x_2,x_3]$, is 
$$\{x_3,x_1+x_2^2\}.$$ 
Thus, $\dim_O(\V(F))\neq 0$. 

Next, let us consider the case $\ell =1$. Let $E=\{x_1+u_{11}x_2+u_{12}x_3\}$. The reduced Gr\"obner basis of $\langle F  \cup E \rangle : \langle x_1,x_2,x_3\rangle^\infty$ w.r.t. $\succ$, in $\C(u_{11},u_{12})[x_1,x_2,x_3]$ is 
$$\{x_3,x_2-u_{11},x_1+u_{11}^2\}.$$ 
Hence, as $\dim_O(\V(F\cup E))=0$, we obtain $\dim_{O}(\V(F))=1$. 
\end{ex}

Algorithm~2-3 can be generalized to parametric cases. The key of the generalized method is to compute comprehensive Gr\"obner systems in $(\C(u)[t])[x]$. We illustrate the method with the following example.

\begin{ex}
Let $f=x_1^3+x_1x_3^2+t_1x_1x_2^3+x_2^3x_3 \in \C[t_1][x_1,x_2,x_3]$ and $F=\{\frac{\partial f}{\partial x_1},\frac{\partial f}{\partial x_2},\frac{\partial f}{\partial x_3}\}$ where $t_1$ is a parameter. Let $\succ$ be the total degree reverse lexicographic term order with $x_1 \succ x_2 \succ x_3$. 
\begin{enumerate}
\item[(i)] Let us consider the case $\ell =0$. A comprehensive Gr\"obner system of $\langle F  \rangle : \langle x_1,x_2,x_3\rangle^\infty$ w.r.t. $\succ$ is 
$$\{(\C\backslash \V(t_1^2+1), \{1\}), (\V(t_1^2+1),\{x_1-t_1x_3,x_2^3+2t_1x_3^2\})\}.
$$
Hence, if the parameter $t_1$ belongs to $\C\backslash \V(t_1^2+1)$, then $\dim_O(\V(F))=0$.
\item[(ii)] Next, let us consider the case $\ell =1$. Let $E=\{x_1+u_{11}x_2+u_{12}x_3\}$. A comprehensive Gr\"obner system of $\langle F \cup E  \rangle : \langle x_1,x_2,x_3\rangle^\infty \subset (\C(u_{11},u_{12})[t_1])[x_1,x_2,x_3]$ w.r.t. $\succ$, on the stratum $\V(t_1^2+1)$, is \\

\noindent 
$\{(\V(t_1^2+1),\{(u_{12}^6+3u_{12}^4+3u_{12}^2+1)x_3+(-2u_{12}^3+6u_{12})u_{11}^3t_1+(-6u_{12}^2+2)u_{11}^3,(-u_{12}^4-2u_{12}^2-1)x_2+(-2u_{12}^2+2)u_{11}^2t_1-4u_{12}u_{11}^2,(u_{12}^6+3u_{12}^4+3u_{12}^2+1)x_1+(-6u_{12}^2+2)u_{11}^3t_1+(2u_{12}^3-6u_{12})u_{11}^3\})\}.$\\

The first polynomial $(u_{12}^6+3u_{12}^4+3u_{12}^2+1)x_3+(-2u_{12}^3+6u_{12})u_{11}^3t_1+(-6u_{12}^2+2)u_{11}^3$ is not zero at the origin $O$. Hence, the local dimension of $\V(F)$ is equal to 1 on the stratum $\V(t_1^2+1)$.
\end{enumerate}
\end{ex}

\section{Comparisons}
Here we give results of the benchmark tests. 
All algorithms in this paper have been implemented in the computer algebra system {\sf Risa/Asir} \cite{NT92}. All tests presented in Table~1, have been performed on a machine  [OS: Windows 10 (64bit), CPU: Intel(R) Core i9-7900 CPU @ 3.30 GHz, RAM: 128 GB] and the computer algebra system {\sf Risa/Asir} version 20150126 \cite{NT92}. The time is given in second (CPU time). In Table~\ref{comp}, ``$<$0.0156'' means it takes less than 0.0156 seconds, and ``$> 3h$'' means it takes more than 3 hours. 

We use the total degree reverse lexicographic term order with $x \succ y \succ z$ (or $x \succ y$) in the benchmark tests. We  use the following 10 polynomials. \\
 \ \\
\hspace{1cm}$f_1=x^3+xz^2+axy^3+y^3z+xy^4$\\
\hspace{1cm}$f_2=x^3y+ay^{15}+bxy^{11}+xy^{12}$\\
\hspace{1cm}$f_3=x^4y+y^8+axy^8+bx^2y^4$\\
\hspace{1cm}$f_4=x^3y+ay^4+y^3+y^8x+by^6$\\
\hspace{1cm}$f_5=x^{4}+yz^5+y^4+ax^4z+y^2z^7+z^4$\\
\hspace{1cm}$f_6=x^5y^3+z^8+axz^8+y^6z+byz^5$\\
\hspace{1cm}$f_7=(x^2y+z^4+y^5)^2+ay^6z^4+y^4z^6$ \\
\hspace{1cm}$f_8=x^{10}+x^5y^3+ay^6+3y^{14}+bx^{10}y^5+xy^{14}$\\
\hspace{1cm}$f_9=x^{5}+yz^4+y^3+ax^5y+bx^2y^{7}+z^{4}$\\
\hspace{1cm}$f_{10}=x^{6}+yz^7+ax^3y^4+y^{10}+x^{2}y^5z^4$\\
 \ \\
where $x, y, z$ are variables and $a, b$ are parameters.

\begin{table}[htb]
\begin{center}
\renewcommand{\arraystretch}{1.5}
\caption{Comparison of the three algorithms}\label{comp}

\begin{tabular}{|c|c|c|c|c|} 
 \hline
Problem  & $F$ & \ \ Algorithm~1 \ \ & \ \ Algorithm~2-1 \ \ &  \ \ Algorithm~2-2 \ \ \\  \hline
1 & $\{f_1,\frac{\partial f_1}{\partial x},\frac{\partial f_1}{\partial y},\frac{\partial f_1}{\partial z}\}$ &  0.0156 &   0.0781      &  0.0156   \\  \hline
2 & $\{f_2,\frac{\partial f_2}{\partial x},\frac{\partial f_2}{\partial y}\}$ &  1.375    &  0.0468     & $<$0.0156   \\  \hline
3 & $\{f_3,\frac{\partial f_3}{\partial x},\frac{\partial f_3}{\partial y}\}$ &  27.94    &   0.7031      & $<$0.0156   \\  \hline
4 & $\{\frac{\partial f_4}{\partial x},\frac{\partial f_4}{\partial y}\}$ &  $>3h$    &   $0.0156$        & $<$0.0156   \\  \hline
5 & $\{f_5,\frac{\partial f_5}{\partial x},\frac{\partial f_5}{\partial y}\}$ &  2.719    &  $>3h$     & 0.0156   \\  \hline
6 & $\{\frac{\partial f_6}{\partial x},\frac{\partial f_6}{\partial y}, \frac{\partial f_6}{\partial z}\}$ &  0.063    &   9.969      & 0.2188   \\  \hline
7 & $\{\frac{\partial f_7}{\partial x},\frac{\partial f_7}{\partial y}, \frac{\partial f_7}{\partial z}\}$ &  0.0156    &   $>3h $     & 226.8   \\  \hline
8 & $\{f_7, \frac{\partial f_7}{\partial x},\frac{\partial f_7}{\partial y}, \frac{\partial f_7}{\partial z}\}$ &  0.0156    &   6.922      & 0.3125   \\  \hline 
9 & $\{\frac{\partial f_8}{\partial x},\frac{\partial f_8}{\partial y}\}$ &  $>3h$    &    0.375       & 0.2669   \\  \hline
10 & $\{f_9,\frac{\partial f_9}{\partial x},\frac{\partial f_9}{\partial y},\frac{\partial f_9}{\partial z}\}$ &  $> 3h$    &    $> 3h$      & 1.391   \\  \hline
11 & $\{f_{10},\frac{\partial f_{10}}{\partial x},\frac{\partial f_{10}}{\partial y},\frac{\partial f_{10}}{\partial z}\}$ &  $>3h$    &   $> 3h$      & 6.188   \\  \hline
 \end{tabular} 
\end{center}
\end{table}

As is evident from Table~\ref{comp}, in Problem 2, 3, 4, 5, 9, 10, 11, Algorithm~2-2 results in better performances in contrast to Algorithm~1 and Algorithm~2-1. In Problem 6, 7, 8, Algorithm~1 results in better performances in contrast to Algorithm~2-1 and Algorithm~2-2. Hence, we cannot say that which one is the best in general. However, as Algorithm~2-2 returns all results within 230 seconds, it is better to utilize  Algorithm~2-2 in general. 

If $\V(F)$ has an irreducible component $V_0=\{O\}$, then the ideal $\langle F \rangle$ can be written as $\langle F \rangle=Q_0 \cap Q_1 \cap \cdots \cap Q_{\nu}$ where $Q_0, Q_1,\ldots,Q_{\nu}$ are distinct primary ideals and $\V(Q_0)=V_0$. Actually, Algorithm~1 computes the ideal $Q_0$ and its dimension. In contrast, Algorithm~2-1 and 2-2 compute the ideal $Q_1 \cap \cdots \cap Q_{\nu}$. Hence, if the structure of $Q_0$ is complicated, then we can expect that the computation cost of Algorithm~2-2 is lower than that of Algorithm~1.\\

In the realm of symbolic computation, the standard basis is regarded as a classical or typical tool to handle ideals in local rings. However, to the best of our knowledge, no effective algorithm for computing standard bases of parametric ideals is known. In order to treat local dimensions for parametric cases, we utilize comprehensive Gr\"obner systems.

To conclude this paper, we emphasize again that even though the problems considered in the present paper are local in nature, the proposed algorithms resolve the problems in polynomial rings and they are free from standard bases and Mora's reduction (Tangent cone algorithm \cite{GP,Mo82}).


\appendix

\section{Ideal quotients with parameters}

Several algorithms for computing a basis of an ideal quotient in a polynomial ring are introduced in some textbooks (cf. \cite{cox04,GP}). As, in general, the algorithms utilize Gr\"obner basis computation, the algorithms can be naturally extended to the parametric cases by utilizing comprehensive Gr\"obner systems (see the appendix of \cite{Nabe15}). 

Here, we briefly describe an efficient algorithm for computing ideal quotients with parameters, that utilizes a comprehensive Gr\"obner system of a {\em module}. \\

Let {\boldmath $e$}$_1=(1,0)$ and {\boldmath $e$}$_2=(0,1)$. Then, $\{\mbox{{\boldmath $e$}$_1$},\mbox{{\boldmath $e$}$_2$}\}$ is a free basis of $(\C[x])^2$. Let $\succ$ be a term order on $\Term(x)$ and $\succ_{\bf m}$ be a POT (position over term) module order on $(\C[x])^2$ with $\mbox{{\boldmath $e$}$_1$} > \mbox{{\boldmath $e$}$_2$}$ and $\succ$. The following theorems are from \cite{GP,Kap18}.

\begin{thm}
Let $f_1,\ldots,f_s,q$ be non-zero polynomials in $\C[x]$. Suppose $F \subset (\C[x])^2$ is a $\C[x]$-module generated by 
$\{f_1\cdot \mbox{{\boldmath $e$}$_1$},f_2\cdot \mbox{{\boldmath $e$}$_1$},\ldots,f_s\cdot \mbox{{\boldmath $e$}$_1$},q\cdot \mbox{{\boldmath $e$}$_1$}-\mbox{{\boldmath $e$}$_2$}\}$ 
 and $G$ is a minimal Gr\"obner basis of $F$ w.r.t. $\succ_{\bf m}$. Set $H=\{h \in \C[x] | h\cdot \mbox{{\boldmath $e$}$_2$}\in G \}$. Then, $\langle f_1,\ldots,f_s \rangle : \langle q  \rangle = \langle H \rangle$.
\end{thm}

There exists algorithms and implementations for computing a comprehensive Gr\"obner system of a given {\em module} with parameters (cf. \cite{Kap18,Na10}). Hence, we are able to obtain a comprehensive Gr\"obner system of an ideal quotient with parameters. 

\begin{thm}\label{iqu}
Let $y=\{y_2,\ldots,y_{r}\}$ be new variables such that $t \cap x= \emptyset$. Let $f_1,\ldots,f_s,q_1,\ldots,q_r$ be non-zero polynomials in $\C[x]$. Set $q=q_1+y_3q_2+\cdots+y_{r}q_r$ and let $G$ be a Gr\"obner basis of the ideal quotient $\langle f_1,\ldots,f_s\rangle : \langle q \rangle$ w.r.t. a block term order such that $y \gg x$ in $\C[y,x]$. Then, $\langle f_1,\ldots,f_s \rangle : \langle q_1,\ldots,q_r  \rangle = \langle G \cap \C[x] \rangle$.

\end{thm}

As we know how to compute a comprehensive Gr\"obner systems of $\langle f_1,\ldots,f_s \rangle : \langle q  \rangle$, Theorem~\ref{iqu} also can be generalized to the parametric cases, too. 

An algorithm for computing a comprehensive Gr\"obner system of an ideal quotient is the following.

\noindent
\hrulefill  \vspace{-1mm}\\
\noindent 
{\bf Algorithm~A} \verb|(ideal quotients with parameters)|\vspace{-3.0mm}\\
\hrulefill\vspace{1mm} \\
\noindent 
{\bf Input:} $f_1,\ldots,f_s,q_1,\ldots,q_r \in \C[t][x]$ ($\forall \bar{t} \in \C^m$, $1 \le \exists i \le s$ s.t. $\sigma_{\bar{t}}(f_i)\neq 0$). \\
\hspace{1.1cm}$\succ$: a block term order with $y \gg x$ on $\Term(x \cup y)$.  \\
\hspace{1.1cm}$\succ_{{\bf m}}$: a POT module order on $(\C[x])^2$ with $\mbox{{\boldmath $e$}$_1$} > \mbox{{\boldmath $e$}$_2$}$ and $\succ$. \\
{\bf Output:} ${\mathcal Q}$: a comprehensive Gr\"obner systems of $\langle f_1,\ldots,f_s\rangle : \langle q_1,\ldots,q_r\rangle$ w.r.t. $\succ$. \\ 
{\bf BEGIN} \\
${\mathcal Q}\gets \emptyset$; \\
$q\gets q_1+y_2q_2+\cdots+y_{r}q_r$; \\
$F \gets \{f_1\cdot \mbox{{\boldmath $e$}$_1$},f_2\cdot \mbox{{\boldmath $e$}$_1$},\ldots,f_s\cdot \mbox{{\boldmath $e$}$_1$},q\cdot \mbox{{\boldmath $e$}$_1$}-\mbox{{\boldmath $e$}$_2$}\}$; \\
${\mathcal G}\gets $ Compute a comprehensive Gr\"obner system of $\langle F \rangle$ w.r.t. $\succ_{{\bf m}}$ in $(\C[t][x])^2$;\\
{\bf while} ${\mathcal G}\neq \emptyset$ {\bf do}\\
 \ \ \ \ Select $(\A,G)$ from ${\mathcal G}$; ${\mathcal G}\gets {\mathcal G}\backslash \{(\A,G)\}$;\\
 \ \ \ \ $H \gets \{h \in \C[t][y,x] | h\cdot \mbox{{\boldmath $e$}$_2$}\in G \}$;\\ 
 \ \ \ \ ${\mathcal Q} \gets {\mathcal Q} \cup \{(\A, H\cap \C[t][x])\}$;\\
{\bf end-while}\\
{\bf return} ${\mathcal Q}$;\\
{\bf END} \vspace{-2.0mm}  \\
\hrulefill \vspace{-1.5mm} \\

Let $I, J$ be ideals in $\C[x]$. Since $\Bigl(I : J \Bigr) : J=I : J^2$,  $I : J^\infty$ can be obtained by utilizing the algorithm above.  
Our implementation for saturation with parameters is given by the following algorithm.

\noindent
\hrulefill  \vspace{-1mm}\\
\noindent 
{\bf Algorithm~B} \verb|(saturation with parameters)|\vspace{-3.0mm}\\
\hrulefill\vspace{1mm} \\
\noindent 
{\bf Input:} $f_1,\ldots,f_s,q_1,\ldots,q_r \in \C[t][x]$ ($\forall \bar{t} \in \C^m$, $1 \le \exists i \le s$ s.t. \ $\sigma_{\bar{t}}(f_i)\neq 0$). \\
\hspace{1.1cm}$\succ$ : a block term order with $y \gg x$ on $\Term(x \cup y)$.  \\
{\bf Output:} ${\mathcal Q}$: a comprehensive Gr\"obner systems of $\langle f_1,\ldots,f_s\rangle : \langle q_1,\ldots,q_r\rangle^\infty$ w.r.t. $\succ$. \\ 
{\bf BEGIN} \\
${\mathcal Q}\gets \emptyset$; \\
${\mathcal G}\gets $ Compute a comprehensive Gr\"obner system of $\langle f_1,\ldots,f_s\rangle : \langle q_1,\ldots,q_r\rangle$ w.r.t. $\succ$;\\
{\bf while} ${\mathcal G}\neq \emptyset$ {\bf do}\\
 \ \ \ \ Select $(\A,G)$ from ${\mathcal G}$; ${\mathcal G}\gets {\mathcal G}\backslash \{(\A,G)\}$;\\
 \ \ \ \ ${\mathcal G}'\gets $ Compute a comprehensive Gr\"obner system of $\langle G \rangle : \langle q_1,\ldots,q_r\rangle$ on $\A$ w.r.t. $\succ$;\\
 \ \ \ \ {\bf while} ${\mathcal G}'\neq \emptyset$ {\bf do}\\
 \ \ \ \ \ \ \ \ \ \ \ Select $(\A',G')$ from ${\mathcal G}'$; ${\mathcal G}'\gets {\mathcal G}'\backslash \{(\A',G')\}$;\\
 \ \ \ \ \ \ \ \  \ \ \  \ \ \ {\bf if} $G=G'$ {\bf do} \\
 \ \ \ \ \ \ \ \ \ \ \  \ \ \  \ \ \  \ \ \ ${\mathcal Q} \gets {\mathcal Q} \cup \{(\A',G')\}$;\\
 \ \ \ \ \ \ \ \  \ \ \  \ \ \ {\bf else}\\
 \ \ \ \ \ \ \ \ \ \ \  \ \ \  \ \ \  \ \ \ ${\mathcal G}\gets {\mathcal G}\cup\{(\A',G')\}$;\\
 \ \ \ \ \ \ \ \  \ \ \  \ \ \ {\bf end-if}\\
 \ \ \ \ {\bf end-while}\\
{\bf end-while} \\
{\bf return} ${\mathcal Q}$;\\
{\bf END} \vspace{-2.0mm}  \\
\hrulefill \vspace{-1.5mm} \\

As $\C[t][x]$ is a Noetherian ring and an algorithm for computing comprehensive Gr\"obner systems always terminates,  Algorithm~B terminates.


\end{document}